\documentclass[doublecol]{epl2} 

\usepackage{epsfig}

\newcommand{\Eqn}[1]{Eq.~(\ref{#1})}

\newcommand{\Fig}[1]{Fig.~\ref{#1}}     
\newcommand{\Figs}[1]{Figs.~\ref{#1}}     

\newcommand{\lW}{\linewidth}

\title{Elastic collapse in disordered isostatic networks}
\author{Cristian F.~Moukarzel\footnote{email address:
    cristian@mda.cinvestav.mx}}

\institute{Depto.\ de F\'\i sica Aplicada, CINVESTAV del IPN,\\
  Av.~Tecnol\'ogico Km 6, 97310 M\'erida, Yucat\'an, M\'exico. }


\abstract{Isostatic networks are minimally rigid and therefore have,
  generically, nonzero elastic moduli. Regular isostatic networks have
  finite moduli in the limit of large sizes.  However, numerical
  simulations show that all elastic moduli of geometrically disordered
  isostatic networks go to zero with system size. This holds true for
  positional as well as for topological disorder.  In most cases,
  elastic moduli decrease as inverse power-laws of system size.  On
  directed isostatic networks, however, of which the square and cubic
  lattices are particular cases, the decrease of the moduli is
  exponential with size.  For these, the observed elastic weakening
  can be quantitatively described in terms of the multiplicative
  growth of stresses with system size, giving rise to bulk and shear
  moduli of order $e^{-bL}$.  The case of sphere packings, which only
  accept compressive contact forces, is considered separately. It is
  argued that these have a finite bulk modulus because of specific
  correlations in contact disorder, introduced by the constraint of
  compressivity.  We discuss why their shear modulus, nevertheless, is
  again zero for large sizes. A quantitative model is proposed that
  describes the numerically measured shear modulus, both as a function
  of the loading angle and system size.  \\
  In all cases, if a density $p>0$ of overconstraints is present, as
  when a packing is deformed by compression or when a glass is outside
  its isostatic composition window, all asymptotic moduli become
  finite. For square networks with periodic boundary conditions, these
  are of order $\sqrt{p}$. For directed networks, elastic moduli are
  of order $e^{-c/p}$, indicating the existence of an ``isostatic
  length scale'' of order $1/p$. }
\begin{document}

\maketitle
Disorder in condensed matter gives rise to interesting physics that
has been the object of intense study for decades~\cite{ZMOD82,ZTPO98}.
Important problems to which much effort is still being devoted are the
vibrations and electronic conduction in disordered
solids~\cite{ZMOD82}.  Static properties like the elastic moduli of
disordered systems, on the other hand, appear to be more readily
accessible to analysis, so one would expect them to hold fewer
surprises.  This expectation turns out to be true only for hyperstatic
networks.  Consider, for example, a triangular spring network whose
sites have been randomly displaced within a circle of radius
$\epsilon$, with repose lengths of all springs adjusted accordingly.
All elastic moduli of these positionally disordered (PD) networks are
finite and size-independent for large sizes, and go continuously to
their homogeneous limit when $\epsilon \to 0$~\cite{MANO08}. Formally,
$Y_L(\epsilon) \approx Y_{\infty} - a \epsilon^2$ for large $L$. A
similar behavior is observed for topologically disordered (TD)
networks obtained e.g.~by locally rewiring a small fraction $\epsilon$
of the links.
\\
However, the simple perturbative picture described above is not
universal. It is only valid for elastic networks that, like the
triangular network, are Statically Indeterminate or
\emph{Hyperstatic}, i.e.~have more springs than strictly required to
be rigid.  Networks which are minimally rigid and do not have
redundant springs are Statically Determinate or
\emph{Isostatic}. Their elastic properties show, as their size
increases, a dramatic disorder-weakening effect that is the subject of
this work. We will argue that all elastic moduli of isostatic networks
go to zero with system size, if these have random geometrical
(positional or topological) disorder.
\\
\begin{figure}[!h]
\centerline{\LARGE \textbf{a)} 
\epsfig{file=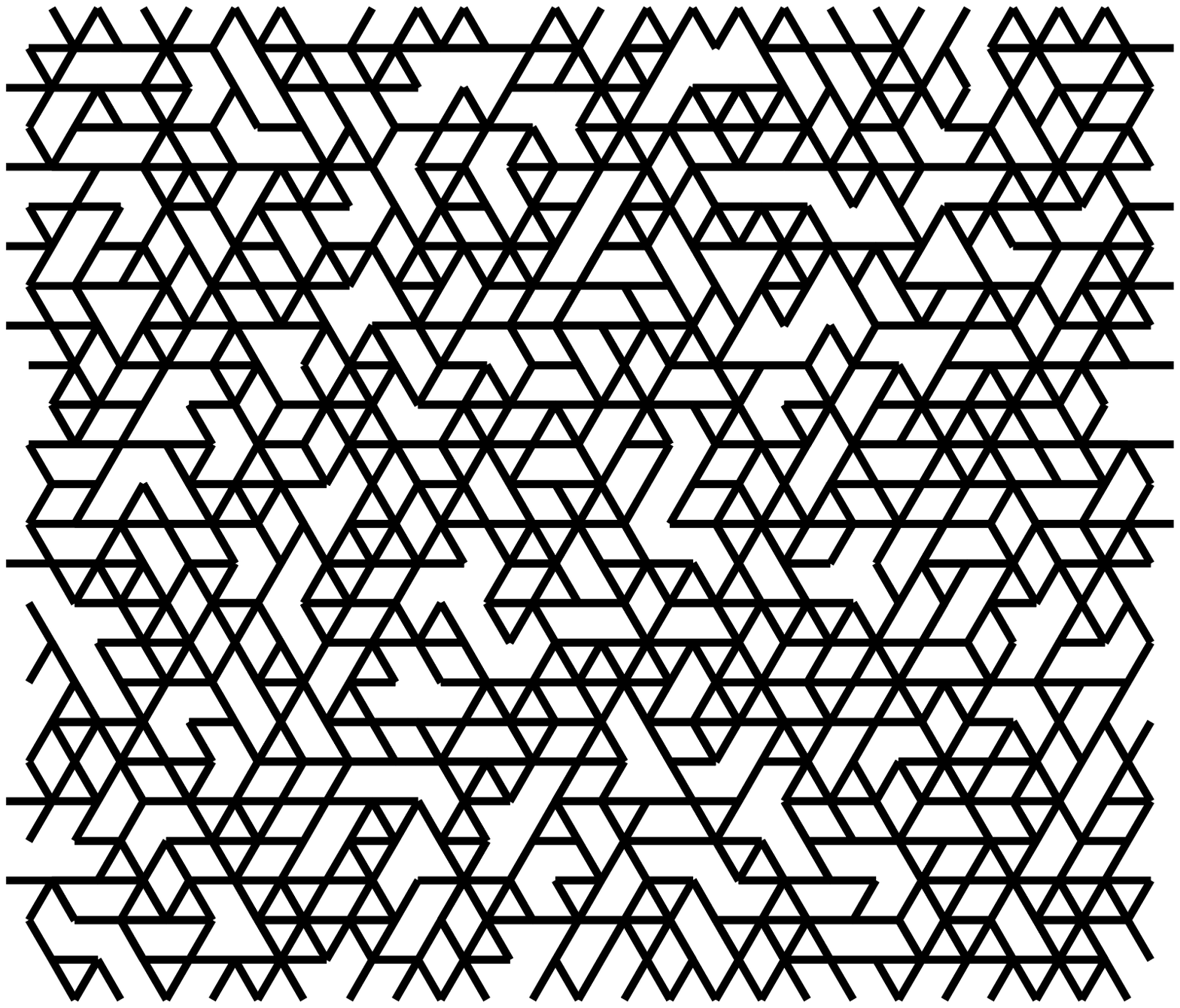,angle=270,width=0.45\lW}
\textbf{b)
\epsfig{file=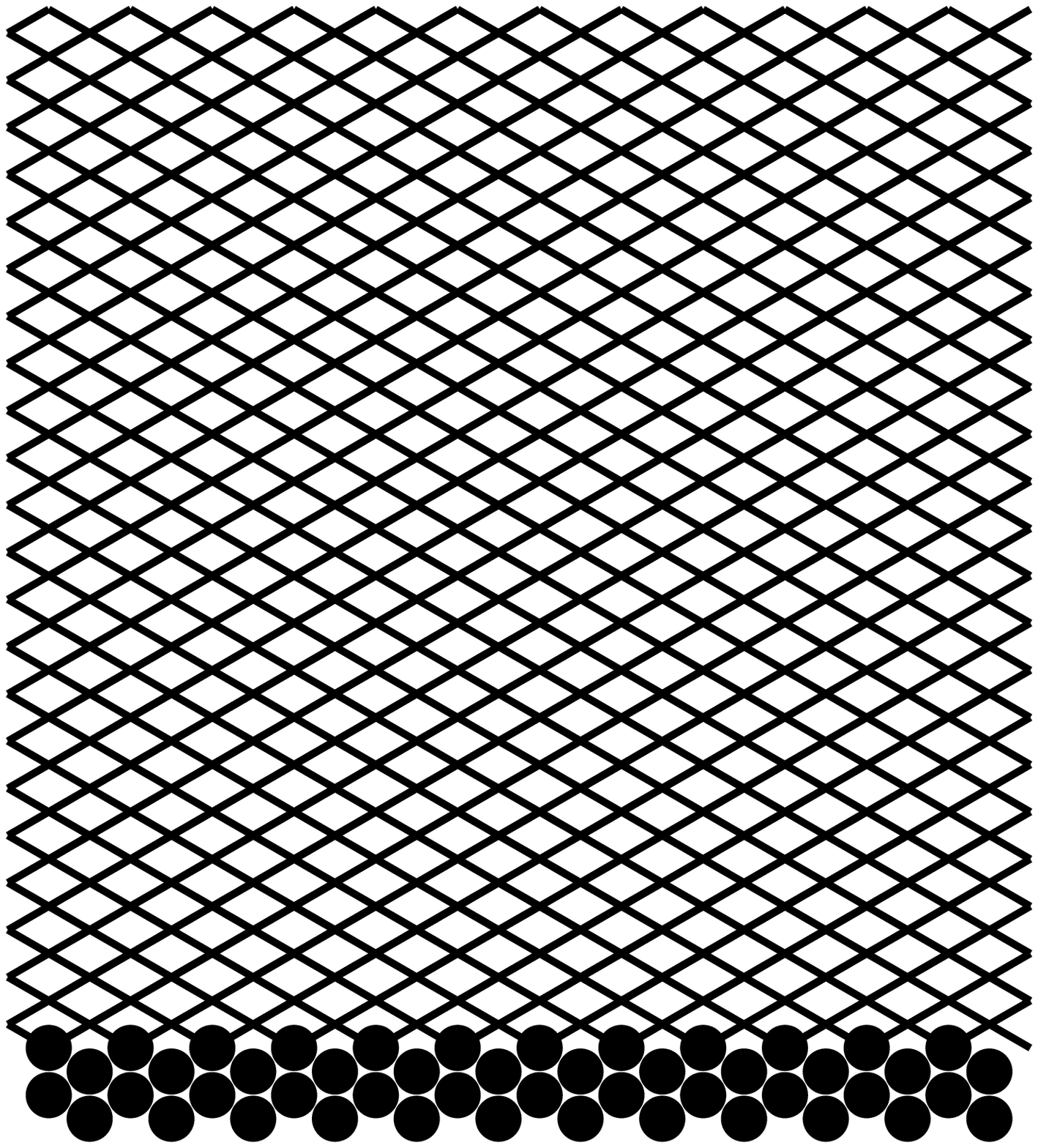,angle=270,width=0.45\lW}
}}
\caption{a) Random isostatic network with periodic boundaries. These
  networks are built by means of a matching algorithm for
  rigidity~\cite{MAEA96}.  b) Positionally disordered square lattice
  attached to a fixed boundary (black circles) and loaded at the top,
  with periodic boundary condition in the horizontal direction.  }
\label{fig:networks}
\end{figure}
An Isostatic Network (IN) is mechanically rigid, and therefore has no
flexibilities (zero-energy modes), other than the trivial
$m(d)=d(d+1)/2$ rigid roto-translations in $d$ dimensions. But the
removal of \emph{any} spring introduces one flexibility, that is, an
isostatic network has no redundant springs.  A simple counting of
degrees of freedom ($d$ degrees of freedom per site) and constraints
(each spring counts as a constraint), originally due to
Maxwell~\cite{MCC64}, shows that a freely-standing isostatic network
of $N$ points in $d$ dimensions has exactly $d N - m(d)$ springs. This
condition is called constraint balance.  Neglecting boundary
contributions, their average coordination is therefore $<z>=2d$.  It
is important to remark that constraint balance is not a sufficient
condition for isostaticity. Additionally, all springs must be
``properly distributed'', in the sense defined in the theory of graph
rigidity.
\\
\begin{figure}[!h]
\centerline{\LARGE \null \hskip 0.1 \lW \textbf{a)} \hskip -0.1 \lW
  \epsfig{file=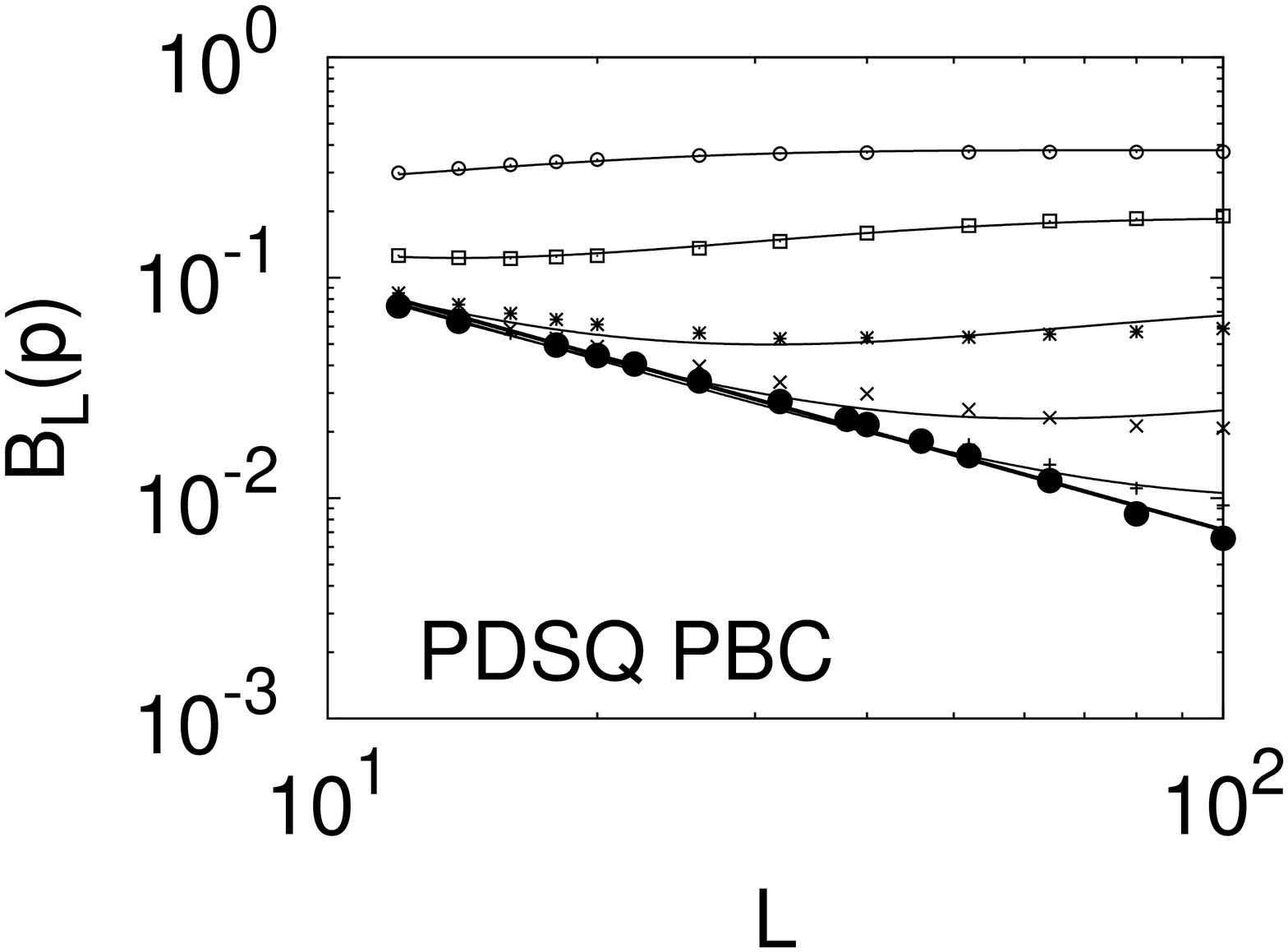,angle=270,width=0.9\lW}
}
\centerline{\LARGE \null \hskip 0.1 \lW \textbf{b)} \hskip -0.1 \lW
  \epsfig{file=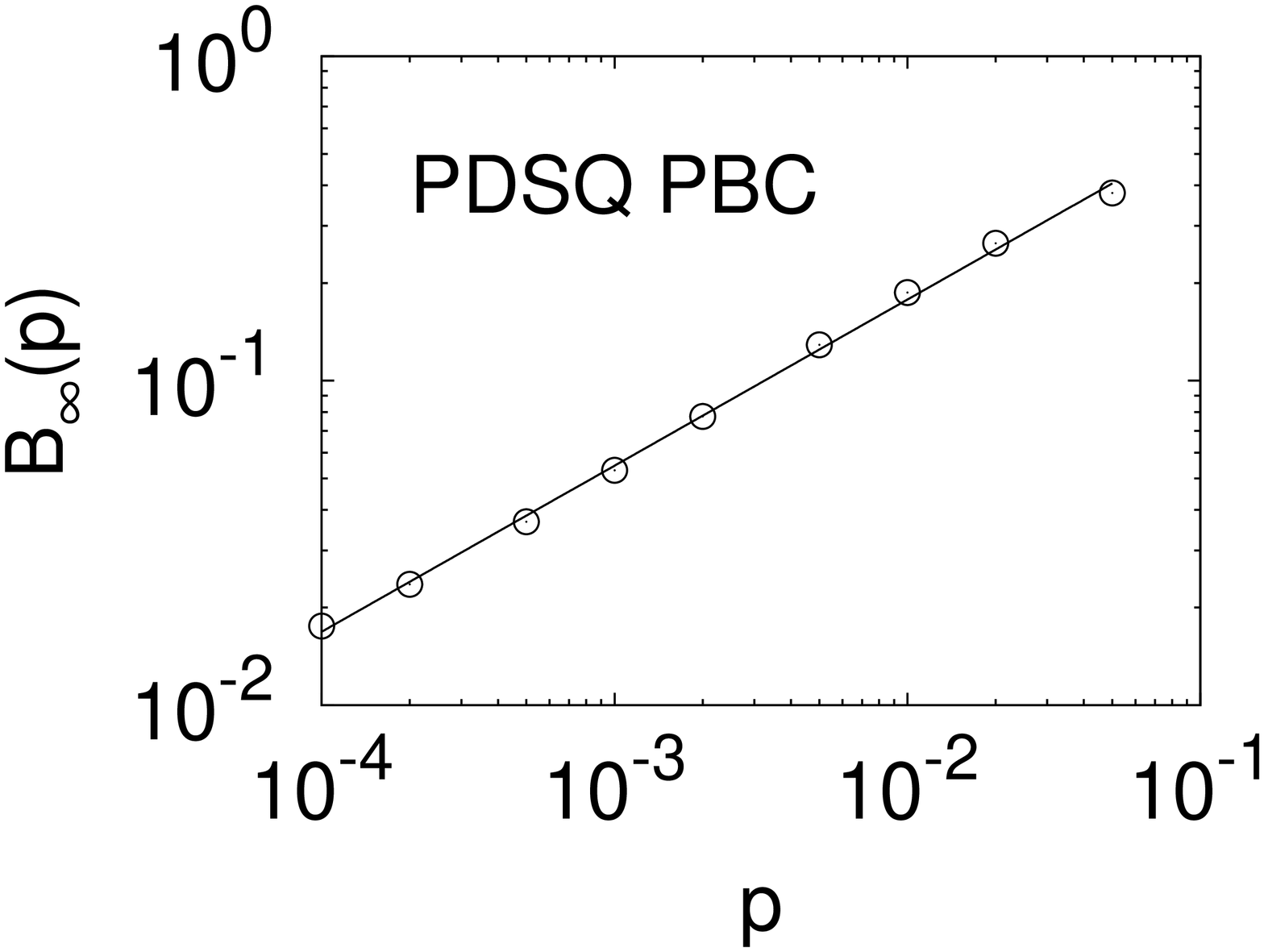,angle=270,width=0.9\lW}
}
\caption{ \textbf{a)} Bulk elastic modulus $B_{L}(p)$ for PDSQ with
  periodic boundary conditions (PBC) in the isostatic case (full
  circles) and with a density $p$ of excess springs equal to $1\times
  10^{-4}$ (plusses), $5\times 10^{-4}$ (crosses), $2\times 10^{-3}$
  (stars), $1\times 10^{-2}$ (squares), and $5\times 10^{-2}$
  (circles). Solid lines are fits to the data (see text). \textbf{b)}
  Asymptotic modulus $B_{\infty}(p)$ vs $p$ for PDSQ with PBC.  The
  full line is a power law fit with exponent $\gamma = 0.53 \pm 0.05$.
  }
\label{fig:emod.sq.pbc}
\end{figure}

\begin{figure}[!h]
\centerline{\LARGE \null \hskip 0.1 \lW \textbf{a)} \hskip -0.1 \lW
  \epsfig{file=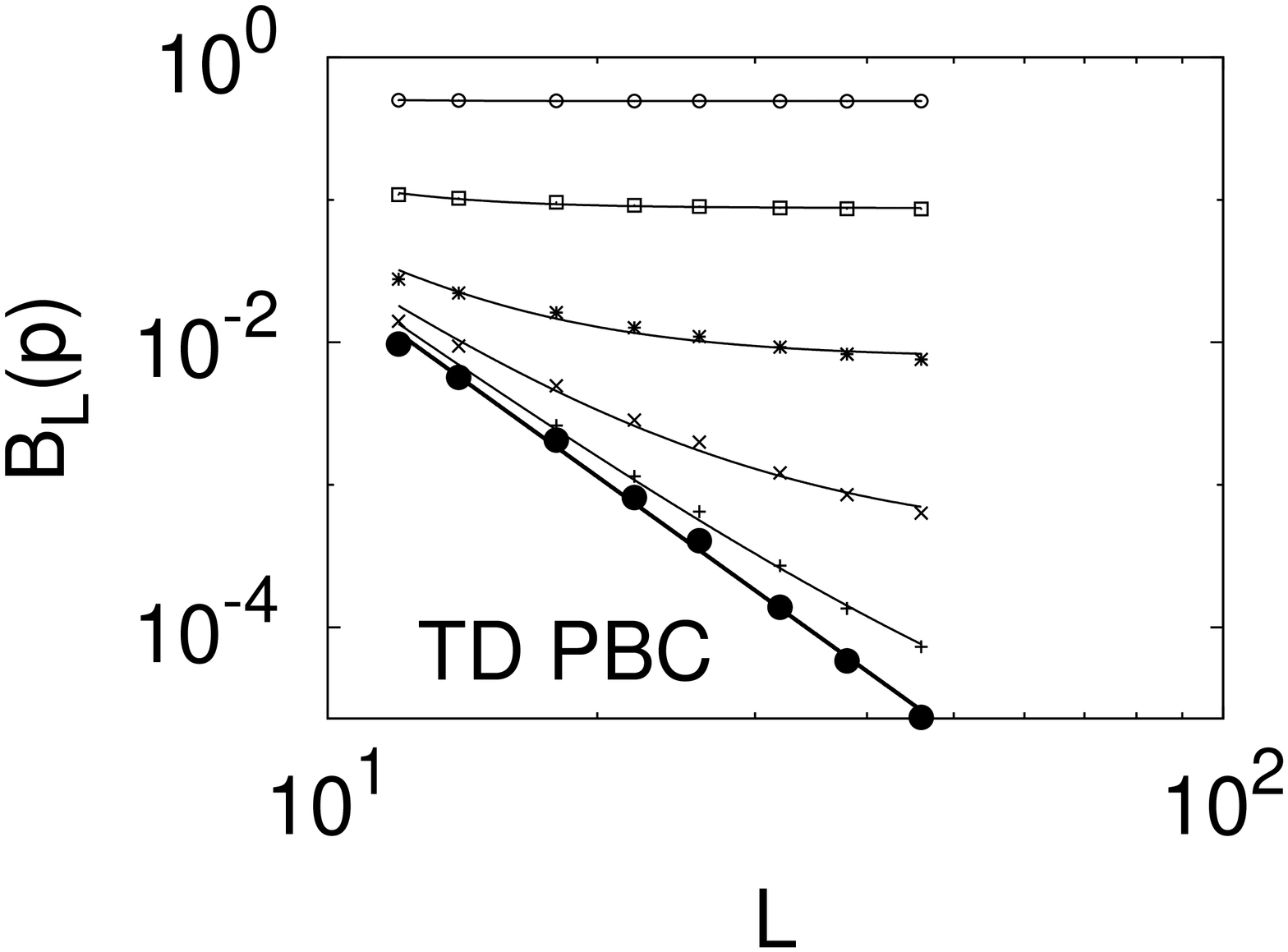,angle=270,width=0.9\lW}
}
\centerline{\LARGE \null \hskip 0.1 \lW \textbf{b)} \hskip -0.1 \lW
  \epsfig{file=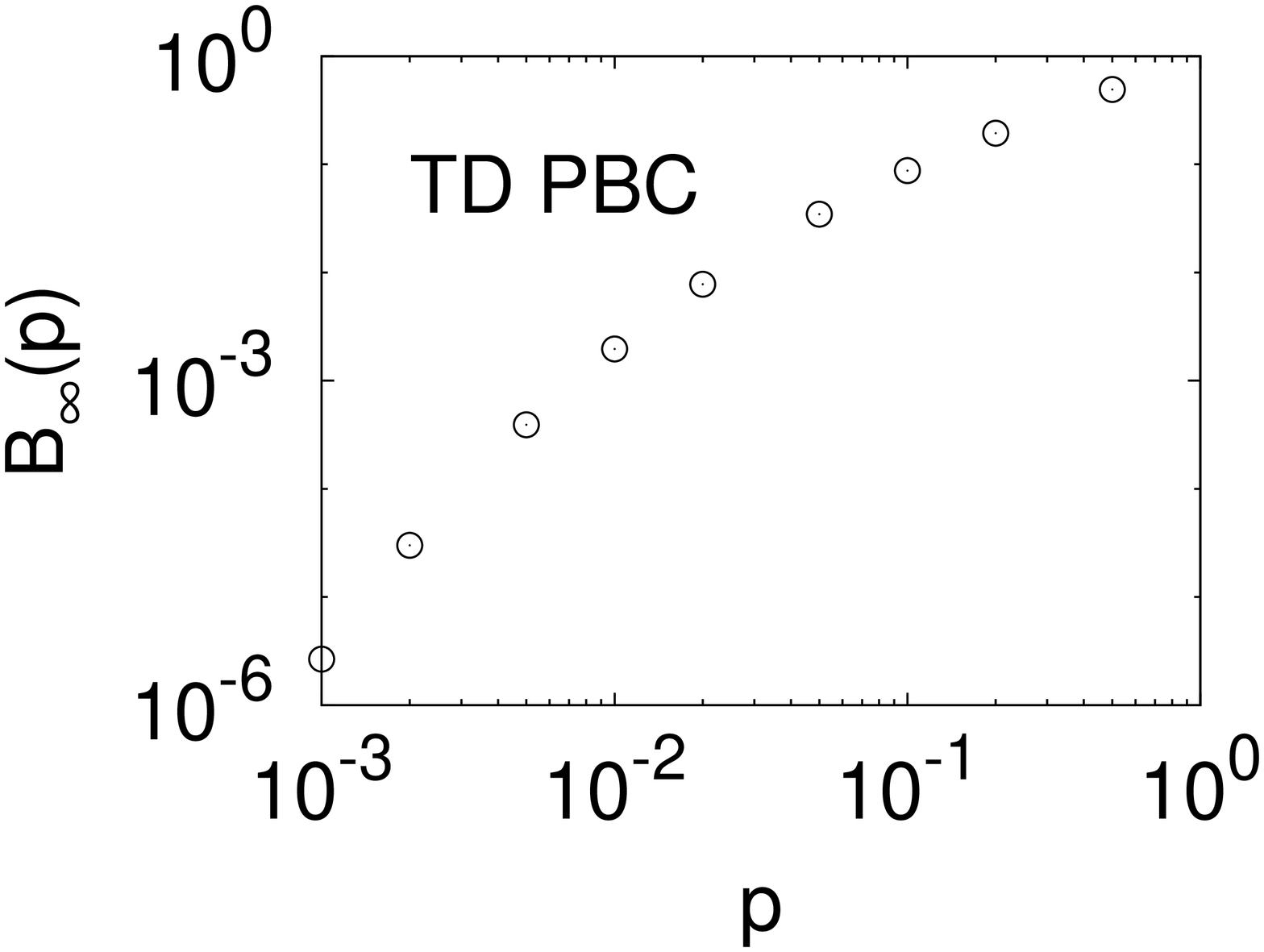,angle=270,width=0.9\lW}
}
  \caption{ \textbf{a)} Bulk elastic modulus $B_{L}(p)$ for TD
    networks with PBC in the isostatic case (full circles) and with a
    density $p$ of excess springs equal to $1\times 10^{-3}$
    (plusses), $5\times 10^{-3}$ (crosses), $2\times 10^{-2}$ (stars),
    $1\times 10^{-1}$ (squares), and $5\times 10^{-1}$ (circles).
    Solid lines are fits (see text). \textbf{b)} Asymptotic modulus
    $B_{\infty}(p)$ vs $p$ for TD networks with PBC.  }
\label{fig:emod.td.pbc}
\end{figure}
\begin{figure}[!h]
  \centering
\centerline{
\epsfig{file=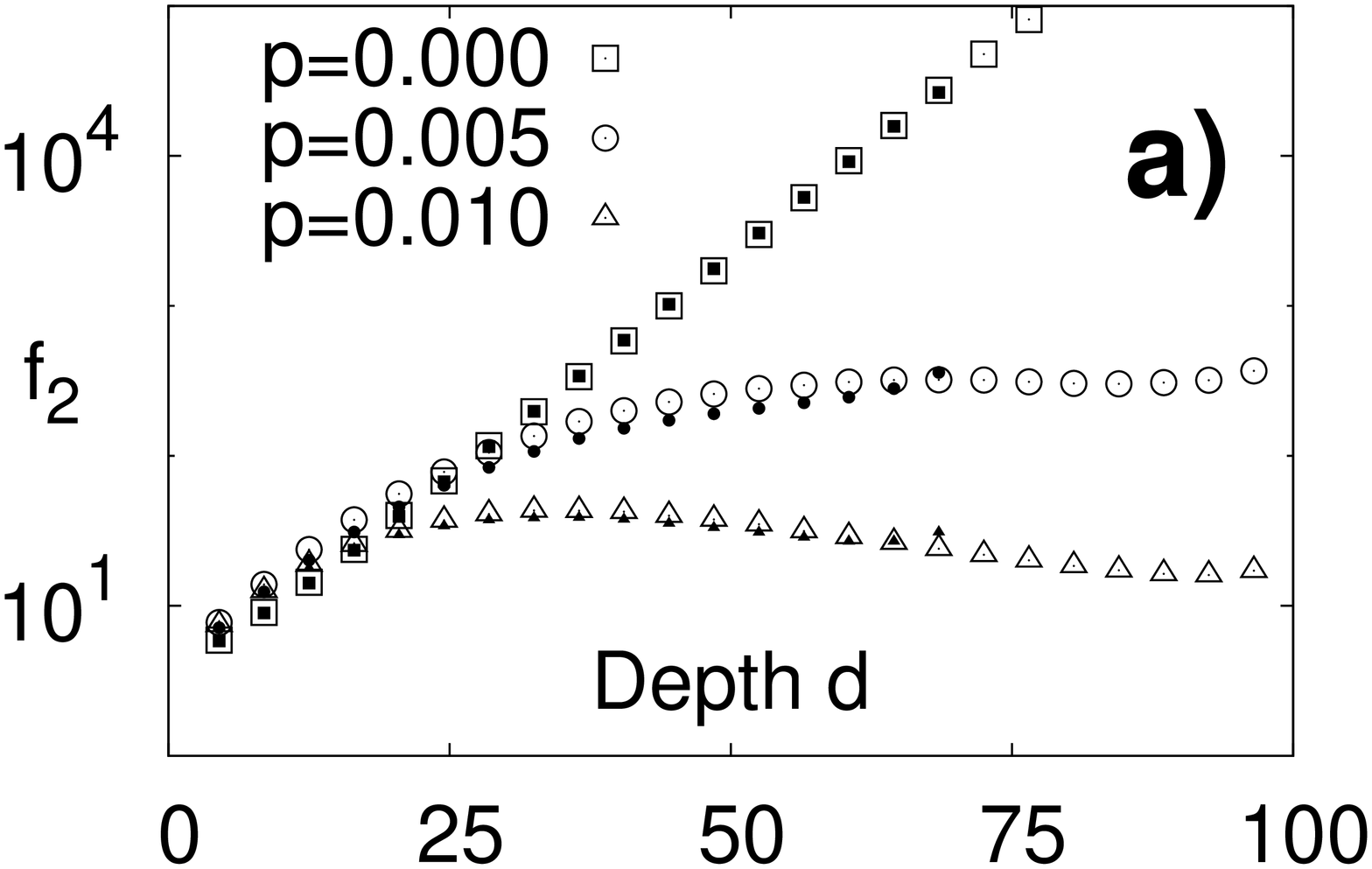,angle=270,width=0.9\lW} 
}
\centerline{
\epsfig{file=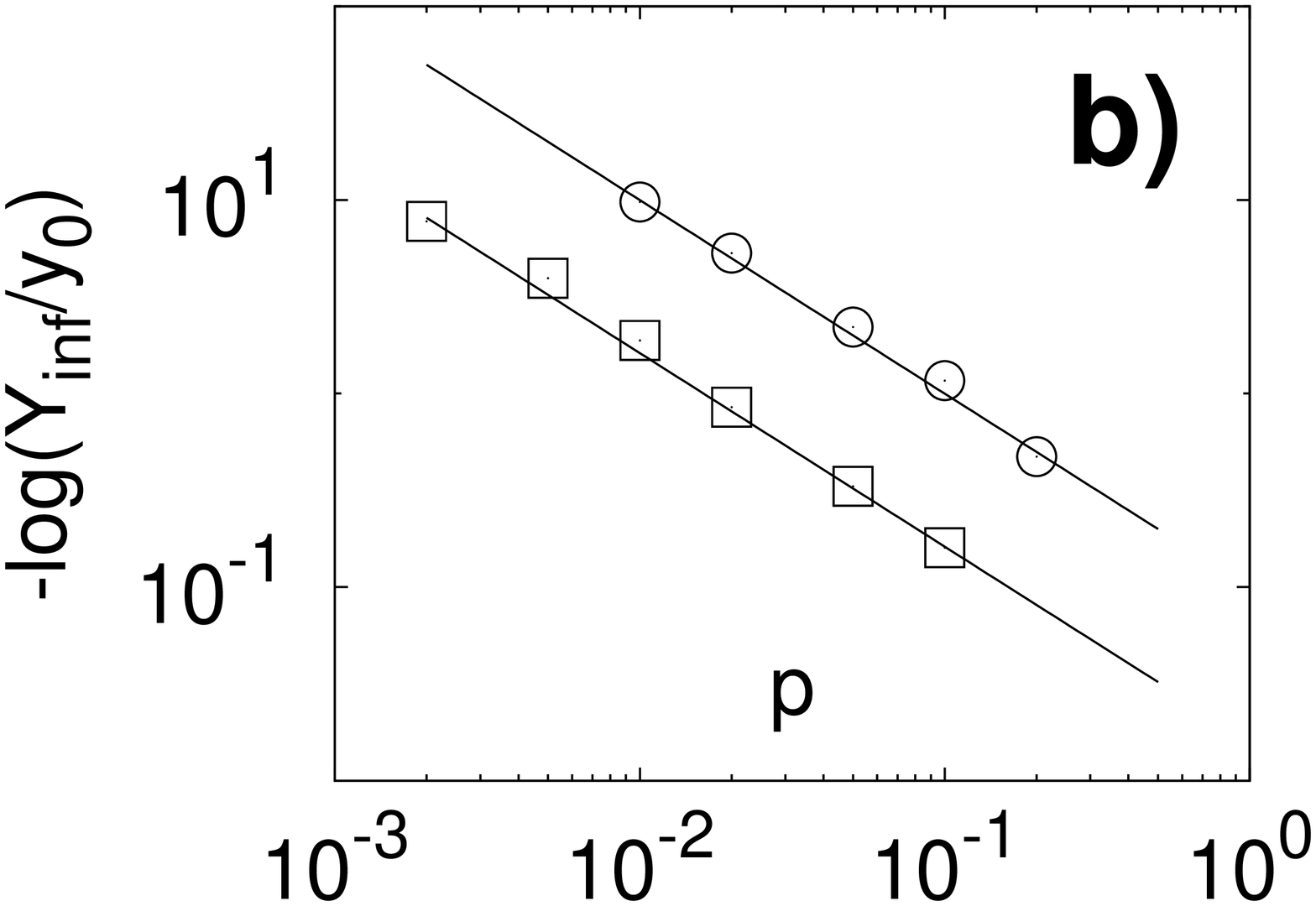,angle=270,width=1.0\lW} 
}
\caption{ a) Second moment $f_2$ of spring stresses at depth $d$ on
  PDDNs with disorder $\epsilon=0.20$, for three different values of
  the density $p$ of excess springs. Data shown are for sizes $L=100$
  (empty symbols) and $70$ (filled symbols). TDDNs show a similar
  behavior. b) Asymptotic compressive modulus $Y_{\infty}(p)$ versus
  overconstraint density $p$, for PDDN (squares) and TDDN
  (circles). Solid lines are fits of the form \hbox{$\log{Y_{\infty}
    /{y_0}}= - c/p$}. }
\label{fig:2}
\end{figure}
\begin{figure}[!h]
  \centering
\centerline{
\epsfig{file=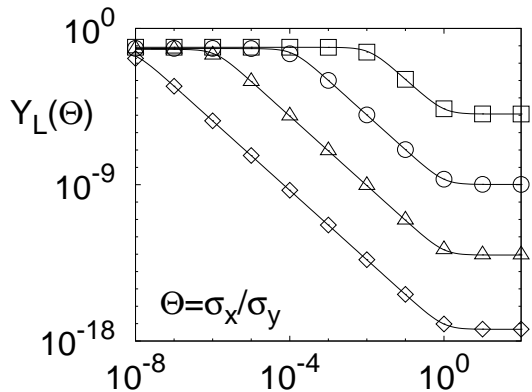,angle=270,width=0.90\lW} 
}
\caption{ Elastic modulus of CSTDDNs, a toy model for compressive
  sphere packings, measured at an angle $\Theta$ from the
  \emph{preparation load}, for sizes $L=100$ (squares), $200$
  (circles), $300$ (triangles), and $400$ (diamonds). Solid lines are
  fits to (\ref{eq:2}). }
\label{fig:3}
\end{figure}

While building topologically disordered isostatic networks is in
general hard (satisfying constraint balance is easy, verifying proper
distribution is highly nontrivial), in two dimensions simple integer
algorithms exist~\cite{MAEA96,JHAAF97} for this task. On the other
hand, topologically regular isostatic lattices are easily
obtained. The simplest examples of such regular INs are the square and
cubic lattices with appropriate boundary conditions. These are
topologically ordered, as every site in the bulk has coordination
$z=2d$.  Given that square and cubic lattices are so common, it is
surprising that their elastic properties in the presence of positional
disorder have not been discussed in detail before.  Furthermore, the
relevance of our findings is not limited to topologically regular
isostatic networks.  In recent years, it has been noticed that several
natural systems self-organize onto topologically disordered structures
that are isostatic, i.e.~minimally rigid without redundant links.
Following the discovery of isostaticity~\cite{MIPT98} in frictionless
sphere-packings (which are models of metallic glasses~\cite{WWLECA99},
colloids~\cite{KKUBI09}, emulsions~\cite{MBWEOC95}, and granular
matter~\cite{MIPT98,OSLJAZ03}), it was later
proposed~\cite{TJCSIN00,CTSAR01} that covalent glasses are isostatic
in a certain composition window.  These further instances of isostatic
networks are topologically disordered, as the coordination number
fluctuates from site to site.  Isostatic systems have a number of
anomalous properties that derive from their being at the verge of
loosing mechanical stability.  Much attention has been devoted in
recent years to the study of the particular
static~\cite{MIPT98,MIIG01,MRMP02,MRMR03,MRFI05,WOTR05,MXLSMA10} and
vibrational~\cite{WOTR05,XWLEVM07,MXLSMA10} properties that isostatic
systems display.
\\
In this work, elastic properties are explored for two different types
of disordered isostatic networks. These are: positionally disordered
square networks (PDSQ), and topologically disordered isostatic
networks (TD).  The latter are built as a random subset of the
triangular lattice, enforcing isostaticity by means of matching
algorithm for rigidity~\cite{MAEA96}. An example of TD is shown in
\Fig{fig:networks}a. For the purpose of these studies, $\epsilon=0.2$
is taken for all disordered networks.  Elastic properties of isostatic
networks are strongly dependent on boundary conditions.  Let us first
consider two-dimensional isostatic networks with periodic boundary
conditions, made of $N=L^2$ sites with $2N$ (properly distributed)
springs~\footnote{ Since two translations are isometries on a torus,
  two of these springs are redundant. Therefore these networks are not
  \emph{strictly} isostatic. However, their number of redundancies is
  not extensive but small, and can be ignored for the purpose of our
  discussion. We will refer to them as being isostatic.}.  The bulk
elastic modulus $B$ of an elastic network with periodic boundary
conditions can be measured as the per-link elastic energy due to an
infinitesimal expansion of all springs. This modulus is finite and
size independent for regular (topologically as well as positionally
ordered) networks as e.g~a regular square lattice.  For PDSQ and TD
networks, on the other hand, there is a disorder-induced elastic
collapse, whereby $B^{\hbox{iso}}_L$ goes to zero as a power law of
system size $L$, as shown in \Figs{fig:emod.sq.pbc}a and
\ref{fig:emod.td.pbc}a. From fits to these data it is found that
\hbox{$B^{\hbox{iso}}_{L} \sim a L^{-\mu}$} with
$\mu^{\hbox{PDSQ}}=1.3 \pm 0.1$, and $\mu^{\hbox{TD}}= 5.5\pm
0.5$. The value of $\mu$ is not universal but dependent on $\epsilon$,
network structure, and boundary conditions. Furthermore, as we later
show, in certain cases the decay of the modulus is not even power-law
but exponential with size.
\\
This disorder-induced elastic collapse only happens for isostatic
networks. If a small density $p>0$ of extra springs per site is now
added~\cite{MDCOR99}, these networks become overconstrained (or
hyperstatic) with average coordination $z=4+2p$. Overconstrained
disordered networks have, in the large-size limit (LSL), a finite bulk
modulus $B_{\infty}(p)$ (See \Figs{fig:emod.sq.pbc}b and
\ref{fig:emod.td.pbc}b). In order to estimate $B_{\infty}(p)$ for
asymptotically large sizes, we fit the bulk modulus data for $p>0$
using ad-hoc expressions, since at present there is no theoretical
prediction for the functional form of finite-size effects.  For PDSQ
networks with $p>0$ we use \hbox{$B_{L}(p)= B^{\hbox{iso}}_{L}
  e^{-c(p) L} + B_{\infty}(p) (1-e^{-c(p) L})$}, with $c(p)$ and
$B_{\infty}(p)$ as fitting parameters for each $p$. From the resulting
fits we find that $c(p) \sim b p^{\sigma}$ with $\sigma \approx
0.5$. This behavior would suggest the existence of a crossover length
of order $p^{-1/2}$. We notice that a characteristic length-scale in
this problem is the average distance between overconstrained links,
and that this length behaves precisely as $p^{-1/2}$ in two
dimensions.  The asymptotic modulus for PDSQ is found to behave as
$B_{\infty}(p) \sim d p^{\gamma}$, with $\gamma = 0.53 \pm 0.05$
(\Fig{fig:emod.sq.pbc}b).
\\
For TD networks, a simple expression \hbox{$B_{L}(p)=
  \{(B^{\hbox{iso}}_{L})^{e} + (B_{\infty}(p))^{e}\}^{1/e}$}, with
$e(p)$ and $B_{\infty}(p)$ as fitting parameters, is found to describe
the data in \Fig{fig:emod.td.pbc}a acceptably well, and $e(p)$ turns
out to be close to one in most cases. Our results for $B_{\infty}(p)$
of TD networks is shown in \Fig{fig:emod.td.pbc}b, where it is seen
that no single power-law of $p$ describes the data well.
\\
We now consider the case of directed INs; those for which rigidity
stems from a boundary, and whose structure is such that all spring
forces can be obtained by local propagation of the loads using force
equilibrium. We find that, while disorder-induced collapse happens for
directed isostatic networks as well, its functional dependence with
system size is entirely different than in the case of periodic
boundaries.  Two examples of disordered directed INs discussed here
are the square positionally disordered directed network (PDDN -- An
example is shown in \Fig{fig:networks}b) and the topologically
disordered directed network (TDDN). TDDN are built by letting each
site be supported by two springs, randomly chosen among its three
lower neighbors on a triangular lattice (For details,
see~\cite{MIPT98,MIIG01,MRMP02,MRMR03,MRFI05}).  Both PDDN and TDDN
are attached to a rigid boundary at the bottom, have periodic boundary
conditions in the horizontal direction, and a free upper boundary
where loads are applied.
\\
\begin{figure}[!h]
  \centering
\epsfig{file=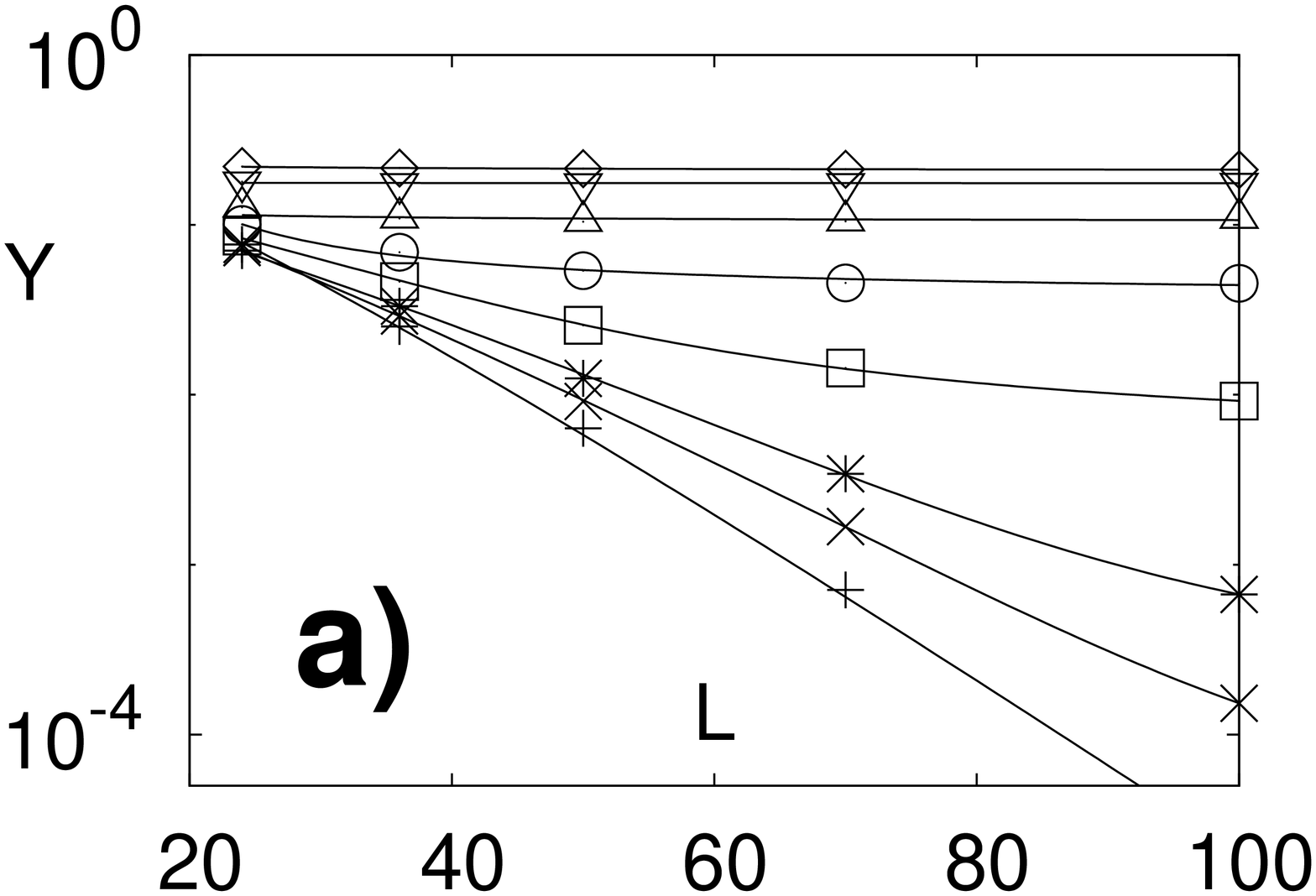,angle=270,width=0.9\lW}
\epsfig{file=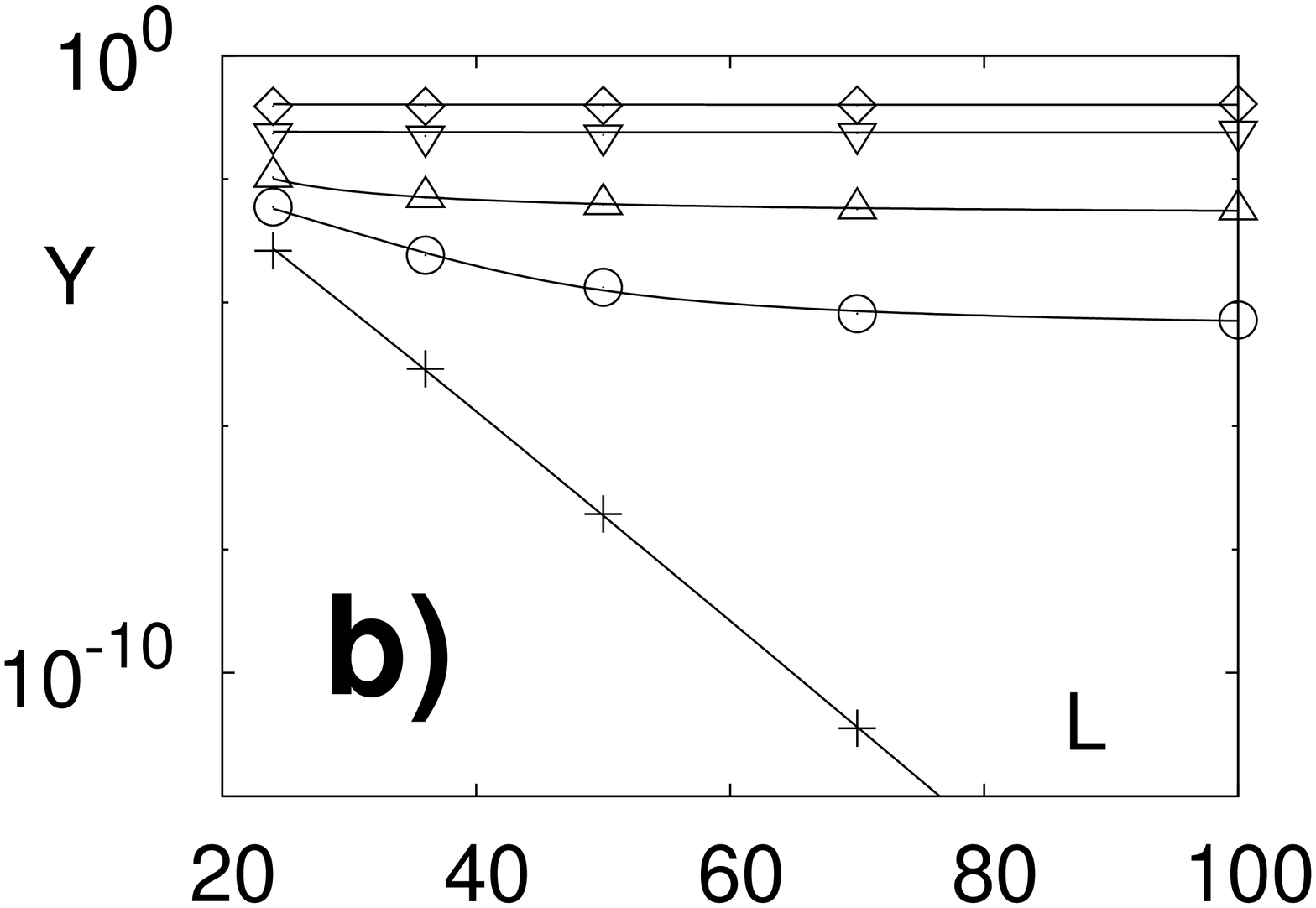,angle=270,width=0.9\lW} 
\caption{ Uniaxial compressive elastic modulus $Y$ for: a) PDDN with
  $\epsilon=0.2$, and b) TDDN, with a density of extra springs $p=0$
  (isostatic case, plusses), $p=10^{-3}$ (crosses), $2\times 10^{-3}$
  (stars), $5\times 10^{-3}$ (squares), $10^{-2}$ (circles), $2 \times
  10^{-2}$ (triangles), $5 \times 10^{-2}$ (downwards triangles), and
  $10^{-1}$ (diamonds). Lines are fits using \Eqn{eq:1}.
}
\label{fig:mod.vs.size}
\end{figure}
Elastic moduli of these networks are proportional to the inverse
per-spring elastic energy resulting from letting a unit load act on
each site of their free upper boundary. If these loads are vertical,
the uniaxial compression modulus $Y$ is obtained, while if the loads
are horizontal, the shear modulus $G$ is measured. In the absence of
disorder, directed square networks have finite uniaxial compression
modulus $Y$ and shear modulus $G$.  For PDDN and TDDN, on the other
hand, both $Y$ and $G$ go to zero exponentially with size.
\Fig{fig:mod.vs.size}a shows the uniaxial compression modulus of PDDN,
while \Fig{fig:mod.vs.size}b shows data for TDDN. The shear modulus
$G$ (not shown) for PDDN and TDDN behaves in a similar way, going to
zero exponentially with size.
\\
Also in the case of directed networks, disorder-induced elastic
collapse is a consequence of isostaticity. Adding a density $p>0$ of
extra springs at each site, overconstrained networks are
obtained. These have, in the LSL, a finite modulus $Y_{\infty}(p)$,
which we estimate by fitting (the use of this expression will be
justified later)
\begin{eqnarray}
  \label{eq:1}
  Y(L,p) =  b L  Y_{\infty}(p) / \ln \left (       
\frac{a+Y_{\infty}(p) e^{bL}}{a+Y_{\infty}(p)} \right )
\end{eqnarray}
to the data in \Figs{fig:mod.vs.size}a,b, with $Y_{\infty}(p)$, $a$,
and $b$ as free parameters when $p>0$. For $p \to 0$, one has
$Y_{\infty} \to 0$ and (\ref{eq:1}) reduces to $Y(L) = a
bL/(e^{bL}-1)$, with only two fitting parameters.
\\
Let us now discuss these observations for directed networks.  For an
elastic network under a load $\sigma$ per unit area, \hbox{$Y \propto
  \sigma^2 / f_2$}~\cite{DRCAVD85}, where $f_2$ is the second moment
of the spring-force distribution. The vanishing of $Y$ with increasing
$L$ then implies $f_2 \to \infty$ in that limit.
\Fig{fig:2}a shows the second moment $f^{(p,L)}_2(d)$ of spring
stresses at a fixed depth $d$ (distance from the top) for several
values of $p$ and for $L=100$ and $70$, on $\epsilon=0.20$ PDDNs.
When $p=0$, $f_2$ grows exponentially with depth as \hbox{$f^{0}_2
  e^{b d}$}.  For overconstrained networks with $p>0$, the growth of
$f_2$ is initially exponential but saturates at a $p$-dependent value
$f^{lim}_2(p)$.  The exponential growth of $f_2$ for $p=0$ can be
understood in terms of previously
reported~\cite{MIPT98,MIIG01,MRMP02,MRMR03,MRFI05} multiplicative
properties of stress propagation on directed INs.  We can now provide
a justification for \Eqn{eq:1}. For PDDN and TDDN, since they are
directed, stresses are determined propagatively from the top down,
without any information at all coming from
below~\cite{MIPT98,MIIG01,MRMP02,MRFI05}.  Therefore $f^{(p,L)}_2(d)$
cannot depend on $L$. This is verified in \Fig{fig:2}a with $p=0$,
where data for $L=100$ and $70$ are seen to behave similarly. When
$p>0$, propagativity no longer holds, as stresses on overconstrained
networks depend on the rheology of the whole system.  Despite this,
\Fig{fig:2}a shows that $f^{(p,L)}_2(d)$ remains approximately
$L$-independent for small $p$ as well, and therefore can be
approximately described by an $L$-independent expression. We choose
\hbox{$f_2(d) \sim ( e^{-b d}/f^{0}_2 + 1/f^{lim}_2 )^{-1}$}, where
$f^{lim}_2(p)$ is the asymptotic value of $f_2$ at large depths. This
simple \emph{ansatz} allows us to calculate $Y^{-1}(L,p) \sim 1/L
\int_0^L dt f_2(t)$ from which \Eqn{eq:1} results, with $Y_{\infty}
\propto 1/f^{lim}_2$. As seen in \Fig{fig:mod.vs.size}a,b, \Eqn{eq:1}
fits our data for PDDN and TDDN quite well.  The resulting asymptotic
values $Y_{\infty}(p)$ for PDDN and TDDN are found to be clearly
incompatible with a power-law form $Y_{\infty}(p) \sim p^{\theta}$,
but they can be fitted quite closely using \hbox{$Y_{\infty} = y_0
  e^{-c/p}$}, as shown in \Fig{fig:2}b.  Writing \hbox{$Y_{\infty}
  \sim 1/f^{lim}_2 \sim e^{-b d_{s}}$}, where $d_{s}(p)$ is a
stress-saturation depth, this last result then implies that $d_{s}
\sim 1/p$.  The following picture then emerges: on directed INs,
stress fluctuations grow
multiplicatively~\cite{MIPT98,MIIG01,MRMP02,MRMR03,MRFI05}, their
scale increases exponentially with depth, and therefore $Y_L \sim
e^{-bL}$. For slightly overconstrained networks, this behavior is
essentially unchanged, i.e.~the system behaves as if isostatic, until
a scale $d_{s} \sim 1/p$ is reached, beyond which $f_2$ saturates and
$Y_L$ attains a finite value $Y_{\infty}$.  We notice that
  an 'isostaticity length' $\ell \sim 1/p$ has been defined in the
  context of vibrational modes of isostatic systems~\cite{WOTR05}. Our
  results show that this same scale determines the elastic properties
  of directed INs. Given that $1/p$ also determines the average
  vertical separation among redundant springs, this might suggest that
  directed isostatic networks behave as if effectively
  one-dimensional, probably due to unidimensional character of stress
  propagation on these systems.  Further work is needed to verify this
  picture in three dimensions. 
\\
We remark that disorder-induced elastic collapse happens for isostatic
networks with any type of local structure and any sort of boundary
conditions, if these have uncorrelated geometric disorder.  Other
forms of randomness, like mass or spring-constant disorder, which
determine the dynamical properties of networks~\cite{ZMOD82}, do not
have an elastic weakening effect on INs.  Additionally, notice that
appropriately chosen correlations in geometrical disorder may
effectively suppress elastic collapse. This is the case with
compressive isostatic packings, i.e.~those built under the constraint
that no contact be subject to traction, as we discuss next. A system
of frictionless stiff spheres self organizes to satisfy this
constraint by rearranging its contacts, for a given external load
(which we call the \emph{preparation load}), by opening contacts
subject to traction, and replacing them with compressive ones. The
resulting isostatic set of contacts is disordered, but not
independently so. Correlations among them have been imposed by the
compressivity condition, under the given preparation load. The absence
of negative forces makes large stresses impossible, as we discuss
later. As a result, the elastic modulus, measured in the direction of
the preparation load, is finite in the LSL. But for any load other
than the preparation load (if further contact rearrangements are
\emph{not} allowed) the elastic response vanishes in the LSL. This has
been observed in numerical studies of compressive sphere packings,
where the bulk modulus is found to be finite for linear elastic
interactions, but the shear modulus vanishes asymptotically, at the
isostatic point~\cite{OSLJAZ03}.
\\
The vanishing of the shear modulus while the compressive modulus is
finite can be illustrated, and discussed quantitatively, with the help
of a toy model for compressive sphere packings, which we call
Compressive-Stress Topologically Disordered Directed Network (CSTDDN).
This model is defined as follows. For a given (e.g. vertical)
\emph{preparation load}, contact forces are calculated by propagation
from the top, layer by layer. At each site, supporting leg
configurations are chosen at random, but only among the two
possibilities (out of three in 2d) which ensure that all contact
forces below this site are compressive
(See~\cite{MIPT98,MIIG01,MRMP02,MRFI05} ).  This compressivity
condition makes the configuration of contacts dependent upon the loads
applied at the top, and also upon the particular set of contacts
chosen at all intermediate steps, thereby introducing correlations.
The reason why these correlated INs have finite modulus in the
direction of the correlating load is simple.  The sum of all vertical
stress components across a given cut must equal the total load applied
at the top, i.e.~contact forces satisfy a ``conservation
equation''. Since negative forces are absent by construction, contact
forces cannot grow arbitrarily large with size as they do on
uncorrelated isostatic networks, where large negative forces can be
balanced by large positive ones so as to still satisfy this
conservation rule.  Therefore the scale of forces remains finite in
the LSL, and so does the compressive modulus $Y$ (\Fig{fig:3}).  But
these correlations are ineffective to limit the scale of contact
forces if now the load at the top differs in any way from the
preparation load, without allowing for further contact rearrangements.
Consider the case of a load that deviates by an angle $\Theta$ from
the preparation load. A CSTDDN, ``prepared'' with a vertical
homogeneous load $\sigma_y$, has a depth-independent second moment of
stresses $f_2^{\sigma_y} = \sigma^2_y/Y_0$ and a finite modulus $Y_o$.
If an additional shear $\sigma_x$ is now applied (without modifying
the network topology), the second moment of the additional stresses
will grow exponentially with depth $x$ as $f^{\sigma_x}_2 \sim
\sigma^2_x a e^{bx}$. By superposition, the resulting modulus $Y \sim
L(\sigma^2_x + \sigma^2_y)/\int_0^L(f^{\sigma_x}_2+f^{\sigma_y}_2) dx$
is
\begin{equation}
  \label{eq:2}
Y = Y_0 \frac{1+\Theta^2 }{1+a Y_0 \Theta^2 (e^{bL}-1)/bl},   
\end{equation}
where $\Theta=\sigma_x/\sigma_y$ measures deviation of the measurement
load from the direction of the preparation load. As seen in
\Fig{fig:3}, our numerical data for the elastic modulus in this model
is described by this expression very well.  Simple analysis of
(\ref{eq:2}) reveals three regimes that are clearly recognizable in
\Fig{fig:3}, where $Y$ behaves respectively as i) $Y \sim Y_0$ for
$\Theta^2 <e^{-bL}/Y_0$, ii) $Y \sim \Theta^2 e^{-bL}$ for
$e^{-bL}/Y_0<\Theta^2 < 1$, and iii) $Y \sim e^{-bL} $ for $\Theta
>1$.
\\
We have thus shown that, while regular isostatic networks have finite
elastic moduli, the elastic constants of geometrically disordered
isostatic networks vanish in the limit of large system sizes, although
in a manner that depends on structure and boundary conditions. INs
with periodic boundary conditions have a bulk elastic modulus that
goes to zero as a power-law of size $L$ with a non-universal exponent.
For the case of directed INs grounded to a rigid bar and loaded at the
opposite end, all moduli go to zero exponentially with size
\\
The vanishing of elastic modulus with size has been previously
reported~\cite{CTSAR01} for self-organized isostatic networks. These
backbones, being the isostatic subset of a random rigidity percolation
backbone~\cite{MDSBA95}, are necessarily fractal. We have shown that
even compact isostatic structures, including the square and cubic
lattices, have vanishing moduli in the large-size-limit if they are
geometrically disordered.
\\
As a closing remark, we note that models of self-organization in
glasses~\cite{TJCSIN00,WBMGSR00} propose that certain calcogenides may
form isostatic structures in order to lower their elastic energy by
avoiding self-stress. This argument should be modified, in the light
of the results reported here, when a large enough external load $F
=\sigma \Omega $ acts on the melt. The total elastic energy per unit
volume is $e_{self} + e_{load}$ where $e_{load} \sim \sigma^2/Y$ is
\emph{maximized} for an isostatic structure, as we have shown.  The
minimal energy structure would then be determined by a load-dependent
balance between two opposing effects: reducing self-stress by making
the structure closer to isostatic, and increasing the elastic modulus
by making the structure more overconstrained.
\acknowledgements
The author thanks M.~Wyart, P.~Duxbury and M.~V.~Chubynsky for
useful discussions.  

\end{document}